\documentclass[titlepage,twoside,12pt]{article}
\usepackage{amssymb}
\usepackage{amsfonts}
\newcommand{\D}{\displaystyle}

\begin{document}

{\Large \bf Mathematics and "The Trouble with Physics", \\ \\ How Deep We Have to Go ?} \\ \\

{\bf Elem\'{e}r E Rosinger} \\
Department of Mathematics \\
and Applied Mathematics \\
University of Pretoria \\
Pretoria \\
0002 South Africa \\
eerosinger@hotmail.com \\ \\

{\bf Abstract} \\

The parts contributed by the author in recent discussions with several physicists and
mathematicians are reviewed, as they have been occasioned by the 2006 book "The Trouble with
Physics", of Lee Smolin. Some of the issues addressed are the possible and not yet
sufficiently explored relationship between modern Mathematics and theoretical Physics, as well
as the way physicists may benefit from becoming more aware of what at present appear to be
certain less than fortunate yet essential differences between modern Mathematics and
theoretical Physics, as far as the significant freedom of introducing new fundamental concepts,
structures and theories in the former is concerned. A number of modern mathematical concepts
and structures are suggested for consideration by physicists, when dealing with foundational
issues in present day theoretical Physics. \\
Since here discussions with several persons are reviewed, certain issues may be brought up
more than one time. For such repetitions the author ask for the kind understanding of the
reader. \\ \\

{\bf 1. For a Genuine Freedom and Creativity in Concepts \\
        \hspace*{0.4cm} in Physics} \\

\begin{quote}

\begin{quote}

\begin{quote}

"The quantum enigma has challenged physicists for eight decades. Is it possible that crucial
clues lie outside the expertise of physicists ? Remarkably, the enigma can be presented
essentially full-blown to non-scientists. Might someone unencumbered by years of training in
the {\it use} of quantum theory have a new insight ?" \\

\hspace*{1cm}                                B Rosenblum, F Kuttner :

\hspace*{1cm}                                Quantum Enigma (p. 13)

\end{quote}

\end{quote}

\end{quote}

The trouble with physicists, to paraphrase the title of  the 2006 book "The Trouble with
Physics" by Lee Smolin, is not so much with the fact that they do not know enough modern
Mathematics, or that instinctively, they do not really understand the role of Mathematics, and
tend to look at it mostly like ... having to go to the dentist ...

Rather, the far more important trouble is that their way of thinking, more precisely, their
repertoire of {\it fundamental concepts} recalls that of the mathematicians prior to Newton. \\

But let us not get ahead of ourselves, quite a few physicists may instantly reply : after all,
most of the mathematical concepts are not exactly of a physical nature, thus they can range
quite freely, well outside the realms relevant to Physics.

Well, quite fortunately, two facts come into play here, and to a significant extent they lay to
rest the possible concerns some physicist may happen to have about some alleged lack of freedom
in introducing radically new fundamental concepts into modern theoretical Physics, concepts
which may, among other possible sources, turn out to have originated in Mathematics. Namely :

\begin{itemize}

\item The term "physical", and in fact, the very discipline of Physics in its wholeness, has
never yet been defined in a clear, comprehensive and definitive manner. And in fact, it could
not, and what is even more important, it should never be defined so. After all, the realms
relevant to Physics are, most likely, as unlimited potentially as are in the case of other
sciences ...

\item Quite a few initially pure mathematical concepts have over the time entered theoretical
Physics as having nothing short of fundamental importance. To mention only two of them :
spaces of higher, or even infinite dimensions, and the complex numbers.

\end{itemize}

Consequently, it would be very hard to argue in favour of any kind of a priori limitations
imposed on fundamental concepts in order to be appropriate for theoretical Physics,
limitations resulting from the requirement that the concepts have an alleged "physical" nature,
and do so according to that anyhow vague qualification, not to mention the particular way it
may happen to be understood at one or another specific time. \\

But to be more to the point, let us recall the fundamental difference between Plato and his
star pupil Aristotle. The latter turned out to be by far the best "quantity surveyor" known in
human history, and produced a most impressive amount of respective reports about Nature,
although lived about two decades less than the former.

Indeed, for Aristotle, the things were given once and for all, given all of them in the realms
of Manifest Creation, and all that was left was to take cognisance of them, classify them, and
try to explain their connections with each other.

Of course, a lot of imagination and thinking - not to mention hard learning - had to go into
that surveying venture. And even more so due to that arrogant habit of ancient Greek thinkers
not to lower themselves to the level of mere experiments, and instead, to try to find out
everything by pure thinking alone, with at most the use of a few basic and simple direct
observations of readily available natural phenomena.

That was how, among others, Aristotle decided that in a horizontal motion of any object force
is proportional with velocity and not acceleration, acceleration which he did not have any
concept of, or that the Earth is not moving, since a stone let to fall freely from the top of
a tower hits the Earth at the foot of that tower ... \\

Well, mathematicians until Newton did not do anything else but endlessly ruminate upon
mathematical concepts and structures, be they geometric or algebraic, which had been known
ever since ancient times. \\

Due to Newton's Calculus, however, and even more so somewhat later, and certainly starting for
instance with Group Theory initiated by Galois in the early 1800s, things changed quite
incredibly in Mathematics. And the awareness, even if not always quite explicit, arouse that,
after all, we were rather at the beginning of our Human History as far as Mathematics was
concerned. Therefore, it was at least as important to start and develop new theories, that is,
introduce fundamentally new mathematical concepts and structures, as it was to keep endlessly
working within the already existing ones ...

And in this manner, modern Mathematics was born, with its rather amazing propensity to
introduce ever new major, and in fact, fundamental realms of mathematical concepts, structures,
theories, and of course, the corresponding thinking and results ...

Suffice in this regard to mention Cantor's Set Theory, introduced in the second half of the
1800s, which became the foundation of just about all of modern Mathematics. Then in the 1940s,
Category Theory was introduced by Eilenberg and Mac Lane, a theory which is yet more
fundamental than that of Cantor ... \\

And it is precisely this incredible freedom of creation of whole new theories, including
fundamental ones, together with their respective structures - all of them based on radically
new concepts - which, so far, is unique to modern Mathematics ... \\

Philosophy has always had a similar, if not in fact, larger freedom, and we can already fully
see it with the pre-Socratic ancient Greek philosophers. However, unlike Mathematics, the
trouble with philosophy is that it has far too much such freedom. Thus it recalls modern
atonal music, while modern Mathematics rather recalls infinitely many systems of harmony, each
of which has certain clear rules of generation or creation, with the fun being precisely to
keep the balance between freedom and rules, a fun so much missing in atonal music ... \\

Of course, just like in Physics and other human endeavours, very few mathematicians are in the
venture of starting new theories by introducing new fundamental concepts.

In fact, the vast majority which is not, tends to downplay that division which Lee Smolin
calls "seers versus craftspeople". Indeed, in modern Mathematics the rather bland terminology
has gained currency which is calling those few  by the name of "theory makers", while all the
rest are called "problem solvers". And obviously that terminology intends to place on some
sort of equal footing the two categories.

However, this division is certainly not a mere matter of a rather unimportant choice, one like
for instance between, say, vanilla and chocolate ice cream ...

On the contrary, it corresponds to an essential distinction, one that no amount of
manipulation based on the brute number of those involved on one of its sides can hide from a
more careful observer. Indeed, those relatively much fewer "theory makers" are far more
important, even if not always in the shortest run.

After all, in philosophy that crucial difference has for ages been very well known and rightly
appreciated. Most certainly, those who merely develop a philosophy originated by someone else
are seldom, if at all, seen as genuine philosophers ...

And quite the same goes on as well in art in general ... \\

And to be still more to the point when it comes to the crucial role of new fundamental
concepts in Physics, let me briefly recall an exchange of letters in mid 1980s with David Bohm
in which I asked him about the following situation, a situation I found - and still find -
highly questionable :

\begin{quote}

How come that the originating heroes of the Copenhagen Interpretation of quantum mechanics
were so immensely, if not in fact arrogantly, proud about having introduced a completely new
way of doing science, while at the same time, they were insisting on the Principle of
Correspondence ? \\

Indeed, how was it possible to claim to have inaugurated such a completely new paradigm in
Physics, and also in science in general, while on the other hand, still be totally nailed to
stone age type fundamental concepts like position, momentum, mass, energy, etc. ... ? \\

After all, in our times, even engineers such as those involved in Systems Theory happen to
have gone beyond stone age type concepts, when introducing such a distinction between physical
entities as described by {\it intensive} versus {\it extensive} ones ...

\end{quote}

David Bohm happened to appreciate my question. However, he never managed to reply in ways
touching more deeply upon the issues involved ... \\

And if we happen to mention Bohm, we can as well recall that the concept of {\it information},
certainly not one of stone age type, is barely making its entrance among the fundamental
concepts in theoretical Physics. \\

Lately, there has in this regard been a lot of talk about "information being physical" ...

However, this inevitably remains quite meaningless, as long as one of the terms, namely,
"physical" is not clearly enough defined. Indeed, there is an obvious asymmetry between the
extent that the two concepts involved are clearly definable. As far as {\it information} is
concerned, its modern meaning, at least since Claude Shannon, is manifestly less vague than
that of {\it physical}, which is merely able to elicit intuitive feelings and possibly vague
ideas which, therefore, may quite likely differ significantly from person to person, not to
mention, from one period of time to another ... \\

Therefore, the ongoing talk about "information being physical" only makes more acute the
challenge to come up with a proper definition of what "physical" is supposed to mean ...

Otherwise, such a talk is a mere attempt to include a modern and fashionable concept like
"information" into what physicists would like to consider as being "physical", yet are not
able or willing to specify in any more clear manner ...

And since a definitive definition of "physical", or for that matter, of Physics as such, is
not only unlikely to emerge, but it may as well be undesirable, it is better to set aside such
meaningless talk as that about "information being physical" ... \\

Recent major interest in Quantum Computation and Quantum Information contributed to the spread
of such rather loose talk about "information being physical".

A possible positive effect may turn out to be an increased awareness of the extent to which
the concept of information is indeed fundamental to modern Physics. However, this inroad among
the fundamental concepts of theoretical Physics which the concept of information may
eventually achieve is so far manifestly indirect at best ... \\

The traditional materialist view in Philosophy got a major shock with Einstein's famous
relation $E = m c^2$, since from brute matter one now could - and in fact, would have to - go
to considering as equally fundamental the more subtle and versatile energy. And yet, the next
expansion, namely, from the stone age type fundamental concepts such as mass and energy to the
inclusion among fundamental physical concepts of information is still in the making ... \\

In this regard, it is amusing to note the following. In the version of Quantum Mechanics often
called Bohmian Mechanics, [3], information plays a fundamental role. Indeed, Bohm's basic
equations are derived from the Schr\"{o}dinger equation in the following rather shockingly
simple and immediate manner. Let \\

$~~~~~~ \psi = R \exp ( i S / h ) $ \\

be the polar representation of the probability amplitude $\psi$ which appears in the
Schr\"{o}dinger equation \\

$~~~~~~ i h \frac{\D \partial}{\D \partial t} \psi =
              - \frac{\D h^2}{\D 2 m} \nabla^2 \psi + V \psi $ \\

where $R$ and $S$ are real numbers. Now, simply by separating the real and imaginary parts,
one obtains the two equations \\

$~~~~~~ \frac{\D \partial S}{\D \partial t} + \frac{\D ( \nabla S )^2}{\D 2 m}
            - \frac{\D h^2}{\D 2 m}\, \frac{\D \nabla^2 R}{\D R} + V = 0 $ \\

$~~~~~~ \frac{\D R^2}{\D \partial t} +
                 \nabla \left ( \frac{\D R^2 \nabla S}{\D m} \right ) = 0 $ \\

And the remarkable fact is that the first above equation is but a classical Hamilton-Jacobi
one, namely \\

$~~~~~~ \frac{\D \partial S}{\D \partial t} + \frac{\D ( \nabla S )^2}{\D 2 m}
                                                           + \widetilde{V} = 0 $ \\

with the potential \\

$~~~~~~ \widetilde{V} = - \frac{\D h^2}{\D 2 m}\, \frac{\D \nabla^2 R}{\D R} + V $ \\

And here Bohm drew attention to the following {\it unprecedented} fact in any of the earlier
basic equations of Physics, namely that the term \\

$~~~~~~ Q = - \frac{\D h^2}{\D 2 m}\, \frac{\D \nabla^2 R}{\D R} $ \\

in the potential $\widetilde{V}$ does {\it not} depend on the magnitude of $R$, and thus of
$\psi$, and instead, can only depend on the {\it shape} of $R$, or correspondingly of $\psi$.
Therefore, in Bohm's view, this term $Q$ - which he called the {\it quantum potential} - is
obviously about the {\it information} content of $R$, and hence of $\psi$. \\

As it happens, this interpretation which for the first time in Physics brings an {\it
essential} and {\it direct} involvement of information into a fundamental equation, and in
this case, into the very foundations of quantum dynamics as described by the  Schr\"{o}dinger
equation, has nevertheless not been accepted widely enough ...

And even if Bohm's specific interpretation is not to be accepted, this still need not
necessarily mean that the utter simplicity and inevitable directness with which the term \\

$~~~~~~ Q = - \frac{\D h^2}{\D 2 m}\, \frac{\D \nabla^2 R}{\D R} $ \\

pops up in the potential $\widetilde{V}$ should so easily be disregarded or dismissed. Indeed,
the presence of such a term which does not depend on the magnitude of one of its constitutive
entities, in this case $R$, that is, $\psi$, is completely unprecedented in such fundamental
equations of Physics as is the case with the Schr\"{o}dinger equation. \\

And yet, most quantum physicists are not at all impressed, and instead, keep holding to
fundamental concepts of such stone age type as position, momentum, mass, energy, etc., ... \\

And then, the effect is that we have some extraordinarily imaginative and creative physicists
who, not realizing any of the above, do nothing else but chase themselves restlessly and
rather arbitrarily within the given and narrow realms - should we call it a cage ? - of mostly
stone age fundamental physical concepts ... \\

Mathematicians, prior to Newton, were doing just about the same ... \\

And in this regard, it may be highly relevant to recall the following. \\

For about two millennia, Geometry had been established as set up by the five axioms of Euclid.
And according to the mentality prevailing during all that time, all what was left was to find
new and new properties within that axiomatic framework, some of such properties rather amusing
and exotic, as for instance the fact that quite a number of special points in an otherwise
arbitrary triangle happen always to be on the same circle ... \\

Well, a few mathematicians happened to be deeply unhappy about Euclid's fifth axiom, namely,
that concerning parallel lines. Indeed, unlike the first four, that fifth axiom seemed to be
less of a direct and simple formulation of an intuitively self-evident truth. Not to mention
that it involved in an essential manner the concept of infinity, thus a concept outside of
one's customary realms of experience. Consequently, some of those few mathematicians tried to
prove it from the first four axioms ... \\

As it turned out in the early 1800s, Euclid's fifth axiom is {\it independent} of the first
four ones.

This discovery, due to Bolyai, Lobachevski, and seemingly Gauss as well, opened up the realms
of Non-Euclidean Geometries, which in about one more century proved to be fundamental in
Einstein's General Relativity. \\

And the moral of the story ?

Well, none of the countless results within Euclid's Geometry obtained over two millennia gave
the slightest hint about Non-Euclidean Geometries.

Instead, it was the totally new idea, or rather, the totally new concept of {\it independence}
of one axiom of some other ones which proved to be so extraordinarily fruitful mathematically,
and soon after, also so fundamental in modern Physics ... \\

Yes, till Newton, mathematicians were arrested within the cage of ancient concepts ...

Is the situation with physicists nowadays so different ?

So different when, for instance, the Principle of Correspondence in Quantum Mechanics is still
made in terms of such ancient fundamental concepts like position, momentum, mass, energy,
etc. ?

And not even information ?

Even if information pops up so simply, immediately and inevitably as in Bohm's version of the
Schr\"{o}dinger equation ? \\

The best of modern mathematicians, on the other hand, have for more than a century by now got
out of their similar conceptual cage ... \\

In this way, the trouble with Physics is that, most likely, physicists have far more
imagination and creativity than their given conceptual cage allows, while a dual, but very
different trouble in Mathematics may be that mathematicians do not really know which to pursue
among the immensity of completely new concepts and theories they can so freely and easily
bring forth, and then choose from ... \\

There is also another utterly regrettable problem with the cage restricted thinking of
physicists. Namely, physicists do in fact seem to consider it as their unique, immense and
incomparable ... God given gift ..., a gift of which all others, including of course
mathematicians, are so manifestly bereft, that they all alone, the physicists, can think in
terms of Physics, even if the Physics of that cage ... \\

And this can lead to rather amusing situations, such as for instance that experienced no less
than twice by Einstein himself. \\

First, when he wanted to go from the Special to the General Relativity, he kept doing nothing
else but trying to find his way through that very same cage restricted physical thinking.

Indeed, he did not much bother instead simply to say to himself :

\begin{quote}

Well, why should the laws of Physics be invariant only with respect to inertial reference
frames, and why should they not be invariant with respect to arbitrary smooth enough
diffeomorphisms as well ?

\end{quote}

Similarly, later, when trying a grand unification, again and again he tried to find his way
exclusively through the very same cage restricted physical thinking ... \\

And as we know so well, in the first instance he succeeded quite wonderfully, while in the
second one, more than two decades of work at the end of his life did not lead him
anywhere ... \\

Well, in Einstein's case, and in view of his rather well known truly blessed personal
character, it was most certainly not out of any arrogance that he insisted on doing it through
that cage restricted physical thinking. Rather, he felt deeply all his life that that way was
the only one he could possibly use given the specific gifts of his own mind, a way which, also,
could possibly give him sufficient confidence in the results obtained ... \\

But then, nowadays, are the reasons of physicists still the very same as it happened with
Einstein when he restricted his fundamental concepts to that cage ?

Or rather, a certain amount of good old fashioned professional arrogance may happen to be
involved as well ? \\
After all, all professionals, including mathematicians of course, do often fall for such
temptations ... \\

Bohm, although originated the equivalent version of the Schr\"{o}dinger equation, the
fundamental equation of Quantum Mechanics, a version in which information appears directly and
for the first time in any such fundamental equation, did not himself seem to find anything
particularly objectionable with the Principle of Correspondence, a principle by which Quantum
Mechanics is so much tied to, if not in fact reduced, to the stone age concepts of Classical
Mechanics ... \\

And if information has still not made it to being one of the most fundamental concepts of
modern Physics, one should not be unduly surprised, since a similarly fundamental and somewhat
older concept appears so far to have the same fate. Smolin in his mentioned book stresses
repeatedly that perhaps the main lesson we should learn from Einstein's General Relativity is
in its {\it background independent} nature.

Classical Mechanics, Special Relativity, Quantum Mechanics, and even much of String or
Super-String Theory take place in an a priori given and fixed space-time background which is
independent of the respective physical processes that occur within it.

Radically opposed to that, in General Relativity there is no - and simply there cannot be -
such a fixed background. Indeed, it is precisely the dynamics of masses and energies which at
each moment determine the structure of space, and do so through the Einstein equations. \\

And yet, as even the case of String Theory shows it, that fundamental idea of background
independence of a physical theory is not yet widely enough accepted ... \\ \\

{\bf 2. Physics May Need Yet More General Ideas \\
        \hspace*{0.4cm} and Concepts ...} \\

At first it may appear strange, and no doubt, even more so to physicists, to consider the
possibility that modern theoretical Physics, on its way out of the cage of stone age
fundamental concepts, may make use of ideas and concepts which appear to be {\it more general}
that those customarily understood as having a so called "physical" nature. \\

In the sequel, we shall present a few such allegedly more general ideas and concepts, and we
can note that, as it happens, their possible relation to modern theoretical Physics cannot so
simply be dismissed out of hand.

And the fact is that mathematical ideas and concepts which at first seemed rather unrelated to
Physics had for a longer time by now proved themselves of fundamental importance in modern
Physics. Two obvious and well known such instances are the higher, and even infinite
dimensional spaces, and of course, the complex numbers. \\

As mentioned in [6, section 1], physicists seem to have an unquestioned and unbreakable trust
in what is called the {\it scaling group} of Dimensional Analysis. And then, just about
everything tends to be seen by them in terms of {\it ratios} ...

Of course, this is all fine {\it locally}, that is, for as long as those ratios are not too
small or too large ...

But is this fact really known ?

And for God's sake, when is a ratio too small, or too large ? \\

Well, when one decides that a ratio A/B is too small, one simply says that A is negligible,
and thus disregards it ...

When on the other hand, that ratio is too large, one never wonders whether there is a trouble
in the respective very way of thinking ... \\

In short, all that unquestioned reliance on the usual scaling recalls nothing else but
reducing things at no matter what scale to a mere {\it local linearisation} ... \\

But to be more to the point, and rather amusingly at that, it quite clearly recalls Marx and
the way he was thinking about economics :

When dealing with such a complex phenomenon like the capitalist economy of the second half of
the 1800s, an economy which was on its way to becoming global, he could, and would, only do
the following thing : single out one or another aspect, consider all other ones constant, and
then note whether that particular singled out aspect would increase or decrease.

Needless to say, he never considered the following two facts : in a complex economy lots and
lots of aspects are simultaneously changing and interacting, and even if only one aspect would
change at a time, its increase, or for that matter decrease, is not going to last for ever,
since the whole system is far from being globally linear. \\

For illustration, let us see how the above unquestioned reliance on the scaling group
manifests itself, for instance, in Special Relativity. \\

A most remarkable fact about Special Relativity is that much of its basics can be done with no
more than school mathematics. And that includes such famous relations like $E = m c^2$. This
certainly can come as a great surprise after all the Calculus involved in Classical Mechanics,
Thermodynamics, or Electro-Magnetism. \\

Amusingly, quite simultaneously with the emergence of Special Relativity, we got more and more
convinced about the atomic nature of matter at micro-scales. Thus it was natural to accept
that the scaling group, which is continuous, would no loner work at such micro-scales. \\

In this way, a first major break occurred in physicists blind faith in scaling ...

It did indeed occur, yet it was not assimilated deeply enough ... \\

Also, it did in no way touch the unquestioned use of scaling at the other end of the scale,
that is, at macro-scales ...

Indeed, we still tend to believe, even if not so consciously, that Special Relativity should
work at absolutely any macro-scales ...

And what is it that such a belief happens to be based on ? \\

Well, the fact is that in Special Relativity the upper limit on velocities shows that scaling
does {\it not} apply universally at arbitrary macro-scales.

Amusingly however, there are no such limitations on mass or energy, or for that matter,
acceleration.

And then, the exception with the case of velocity is simply disregarding when it comes to the
unshakable faith in the validity of scaling ... \\

One possible way out of this fixation with scaling, a way quite well understood and
successfully used in modern Mathematics, is offered by {\it non-Archimedean} structures,
[8,9,16,17]. Comments in this regard are presented in the sequel. \\

Amusingly, the failure of usual scaling at macro-scales already happens in the classical
Newtonian framework. In this regard it is quite delightful to read in Smolin's mentioned book,
on page 211, about MOND, that is, Modified Newtonian Mechanics. \\

Well, one had earlier encountered troubles with Newtonian Mechanics, like for instance the
classical one which already happens in our own backyard, namely, with the precession of the
perihelion of Mercury. And this is quite a trouble since the velocities involved are nowhere
near the speed of light. Also the masses involved are not really so large, say, on a galactic
scale, even in the case of the Sun. \\

On top of that, ever since the modern atomic theory, and specifically, Quantum Mechanics, we
all have known that Newtonian Mechanics does not apply at micro-scales. \\

Yet no one seemed to worry that, similarly, it may not apply at macro-scales either ... \\

And this is precisely what MOND proposes. And as it happens, MOND seems to work very well
within galaxies, even if it does not do so outside of them. \\

However, there are plenty of possible alternatives, beyond simply declaring Newtonian
Mechanics, as MOND does, to be wrong on large enough macro-scales. Let us therefore look into
some of them. \\

Aristotle, without of course writing even one single equation, claimed that force is
proportional with velocity, when moving an object horizontally. That would however lead to a
first order ordinary differential equation in displacement. And then, we could not impose two
independent initial conditions, but only one. And this is clearly contrary to the most
elementary everyday experience.

Newton decided that, instead, force is proportional with the velocity of velocity, that is,
with acceleration. And this gives a second order ordinary differential equation in
displacement, thus we can - and in fact must - impose two independent initial conditions.

However, the way from displacement to velocity is given by the operation of derivative, just
like the way from velocity to acceleration. And then, two questions arise :

\begin{quote}

1. Should acceleration be indeed the velocity of velocity, or rather, it should be obtained
from displacement in some other way than the mere second iterate of derivative ? \\

2. What are the assumptions involved in the definition of the derivative ?

\end{quote}

Question 1 is really tough to answer, since it opens up an immense realm of possible
Mathematics, without giving much hint about an appropriate choice. And then instead, let us
focus on question 2. \\

Before that, however, let me mention briefly a problem which has concerned me for a long
time :

\begin{quote}

Position is relative to a reference system, velocity is similarly relative, while acceleration
is absolute in the sense that one can observe it independent of any reference system. \\
Yet velocity is the derivative of position or displacement, and acceleration is the derivative
of velocity. \\
So then, how can two successive derivatives take one form the relative position to the
absolute acceleration ? \\
Or even more funnily, how can a derivative take one from a relative velocity to an absolute
acceleration ? \\
While at the same time, a derivative still takes us from a relative displacement to an equally
relative velocity ?

\end{quote}

And now let us return to question 2 above. Well, two obvious ingredients enter in the
definition of the usual derivative in undergraduate Calculus :

\begin{quote}

2.1. The natural "gauge theory" of the real line R, that is, its commutative group structure
given by the usual addition. And this structure, as a Lie group, is well known to be
unique. \\

2.2. The possibility to go to the limit, namely, any increment being able to tend to zero.

\end{quote}

Obviously, in view of the atomic structure of matter, 2.2. is physically nonsense. Yet so
amusingly, we can use it not only in Newtonian or Einsteinian Mechanics, but also in the
Schr\"{o}dinger equation ... \\

As for 2.1., this leads uniquely to the corresponding multiplication, thus also division of
the usual real numbers. In particular, it leads to the ratios whose limits are the derivatives.

Now the most amusing feature of 2.1. is that it imposes both the local and global structure of
the usual real line $\mathbb{R}$. And as such, it is Archimedean. As for the local structure,
and as mentioned, it is conflicting with the present atomic view of matter, yet it works, more
precisely, no one cares about that conflict when using the derivative operator. \\

Concerning the possible mismatch of the global structure implied by 2.1., so far no one has
ever come up with any complaint, not physicists, not mathematicians ... \\

And then, for the first time, as far as I happen to know, it is precisely with MOND that we
have to ask ourselves the question :

\begin{quote}

Is it Newtonian Mechanics which is wrong at large scales, or rather, is it in fact the very
Mathematics of Calculus which is wrong ?

\end{quote}

And to tell honestly, I am really concerned that - regardless of Newtonian Mechanics being
wrong at large scales - there is something wrong with Mathematics as well at such scales,
wrong at least as far as the {\it assumptions} upon which Calculus is based are concerned.

Yes, in this regard, and at least for me, MOND is a sign that Calculus is wrong not only at
the "micro end" where it still appears to work, but also for the first time at the "macro
end" as well ... \\

And then, what may be an immediate proposal in this regard ? \\

Very simple : instead of the usual "gauge theories" based on groups, one should base oneself
on the far larger class of {\it semigroups}. Of course, here at first one should only talk
about replacing the usual additive and commutative "gauge theory" of real line $\mathbb{R}$
with a suitable semigroup one. \\

And why ? \\

Well, in [18], a semigroup theory, based on credible mathematical reasons was started. That
theory is general, and not only for the real line $\mathbb{R}$. However, related to 2.1. above,
one could, of course, also consider it in the particular case of the real line
$\mathbb{R}$. \\

Recently, concerns about the customary scaling were presented in [6, section 1]. In this
regard, the story with MOND only comes to give a further reason for such concerns ... \\ \\

{\bf 3. Time, What Is Time ?} \\

In good old, time honoured style, Smolin in his mentioned book left the ... desert ... last,
that is, on pages 256-258, which end his presentation and critical remarks about the state of
the art fundamental theories of Physics.

And needless to say, this ... desert ... is nothing else but the issue of {\it TIME} ... \\

There are of course major differences between space and time, as even our simplest, commonest
and most frequent experiences show it so clearly. For instance, we can easily move back and
forth in space, but not in time. Also, we can stop in a place, but not in a moment of time.

Furthermore, there are similarly sharp differences between mass and time. Yes indeed, we can
cut mass up into pieces, we can put pieces of mass together, and we can do all of that pretty
arbitrarily, but we cannot do any of that with time. \\

So that in terms of L, M and T, namely, length, mass and time, which are the three dimensions
in Classical Physics, we can indeed assume L and M each to be described by the real line
$\mathbb{R}$ with its usual commutative group structure given by addition. \\

However, why should we assume the very same about T, when we cannot so easily go back in time,
and when we face in Physics lots of irreversible processes ? \\

Thus for T it would much more likely be the case to be described by a {\it semigroup}, and not
by a group such as given by the additive structure of $\mathbb{R}$.

And in fact, even the commutativity of that semigroup may be questioned ...

And why not ? \\

But the most funny thing is that it is here, namely, precisely with time, where the
Archimedean structure assumed by Descartes and Galileo should be questioned. And if we at last
consider for T a non-Archimedean structure, then as shown in [8,9,16,17], it is most tempting
to abandon as well the one-dimensionality of time. \\

Indeed, it appears to be most natural to consider multi-dimensional, or shall we say ... {\it
FAT time} ...

That is, a time which is not just that ... slim ... one dimensional real line
$\mathbb{R}$ ... \\

In [1], a convincing argument is presented about the need for a radical reconsideration of the
traditional concept of time. Regrettably however, what is suggested is far too vague to start
something with it, something which may have the nature to be falsifiable ... \\

And last, but by no means least, the issue of time had in fact for a long long time been in
the mind of some of the most notable thinkers. Plotinus, the celebrated neo-Platonist, was in
the late 300s wondering about our understanding of time. Less than two centuries later, St.
Augustin wrote in this regard "So what is time ? If no one asks me, I know; if I seek to
explain it, I do not." \\

But then, St. Augustin is also credited with the remarkable, and seldom considered statement
"I know not what I know not ..." \\

And ever since, do we really know that we do not quite know what time is ? \\

In this way, as far as one can understand, one of the most fundamental differences between
usual humans and those who in the East, for instance, are called enlightened is in the
immensely different and more rich perception of time of the latter ... \\

Yet, we do not have to travel all the way to some mythical East in search of certain alleged
to be enlightened gurus. Indeed, ever since Einstein's Special Relativity, that is, for more
than a century by now, we are supposed to know perfectly well about the {\it relativity of
time}, a relativity with respect to the reference frame of any given observer. And this
relativity is so {\it fundamental} that, as is well known, even the {\it simultaneity} of two
events is relative ... \\ \\

{\bf 4. The Egyptian-Archimedean Captivity ...} \\

Since many physicists seem to be less familiar with the distinction between Archimedean and
non-Archimedean structures, a few related details are mentioned here. Further details can be
found in [8,9,16], and in particular in [17, Appendix 1], where the precise respective
definition is presented. \\

For starters, let us mention that umbers alone, that is, each single one of them, be they real
or complex, for instance, are neither Archimedean, nor non-Archimedean. It is instead their
respective {\it totality} as a space, when endowed with a certain kind of order relation, like
for instance, the set $\mathbb{R}$ of all real numbers with its usual order, that the
resulting totality of numbers can turn out to be Archimedean, or on the contrary,
non-Archimedean. And for instance, $\mathbb{R}$ with its usual order structure is Archimedean.

This means that taking for instance as unit the usual number 1, and adding this unit to itself
a sufficiently large but {\it finite} number of times, one can obtain a number which is larger
than any prior given positive real number. In other words, $\mathbb{R}$ is but "one single
walkable world", [16,17], since someone with the step size 1, can reach any point in it in a
sufficiently large finite number of steps. \\

Of course, those in ancient Egypt who are assumed to have started Geometry, could only use
such a mathematical structure. Thus we got stuck with it ever since, and we simply take it for
granted, without being in the least aware that we do in fact make a particular choice, namely,
the Archimedean one. \\

Now on the other hand, ever since 1966, we have a wonderfully developed and most useful
extension of the usual real line $\mathbb{R}$, namely, given by the nonstandard real line
$^*\mathbb{R}$. And $\mathbb{R}$ is but a small subset of $^*\mathbb{R}$. Furthermore,
$^*\mathbb{R}$, just like $\mathbb{R}$ itself, is one dimensional, as well as a field, that is,
one can perform in it all the usual algebraic operations, except of course for division by
zero. \\

The point is that, unlike $\mathbb{R}$, this extended $^*\mathbb{R}$ is non-Archimedean. And
it is so precisely, since it is no longer "one single walkable world". In other words, there
are uncountably many infinite realms, and two different ones {\it cannot} be reached from one
another by a finite number of steps. In addition, there are as well uncountably many
infinitesimal realms. \\

One effect is that one can nicely compute with all sorts of "infinities", of which there is a
whole uncountable range. Consequently, much of the concerns related to various {\it
renormalizations} in Physics may simply go away with the use of the nonstandard real numbers
$^*\mathbb{R}$. And that would happen precisely due to the non-Archimedean structure of the
nonstandard real line $^*\mathbb{R}$.\\

As for the meaning of such nonstandard "infinities", one can simply note the following. When
we employ the usual real numbers in $\mathbb{R}$, a real number, say, $x \in \mathbb{R}$, is
supposed to describe both a {\it position}, that is, a point on the real line $\mathbb{R}$, as
well as a {\it relationship}, or {\it ratio} between two other real numbers, say, $a$ and $b$,
such as for instance expressed in $x = a / b$, where one assumes of course that $b \neq 0$. \\

However, the Archimedean structure of the usual real line $\mathbb{R}$ constitutes nothing
more than "one single walkable world", [16,17]. Thus everything which may happen to fall
outside of it can only be described by some "infinity". Indeed, such an "infinity" is of
course {\it not} one of the usual real numbers $x \in \mathbb{R}$, and therefore, it cannot
indicate a position on the usual real line $\mathbb{R}$, nor can it indicate a usual
relationship or ratio between two usual real numbers.

In this way, an "infinity" can - and should - be seen as nothing else but an indication that
we have in fact {\it gotten out} from the confines of that "one single walkable world" which
has been imposed upon us by the Archimedean structure of the usual real line $\mathbb{R}$. \\

However, when arrested within the Archimedean paradigm, the trouble is three fold :

\begin{itemize}

\item There is nothing, and there cannot within that paradigm be anything outside of that "one
single walkable world". Thus when we hit upon some "infinity", we cannot properly do anything
with that message, since for us it cannot belong to any mathematical structure, thus it cannot
indicate any position, relationship or ratio, and do so in a manner allowing useful
mathematical operations.

\item We miss on the whole of the immensely rich and complex non-Archimedean structure outside
and beyond that "one single walkable world" of the usual real line $\mathbb{R}$. Thus we miss
on the corresponding sophisticated information about the uncountably many different kind of
"infinities", not to mention all the usual algebraic operations which are available with them
within the non-Archimedean context. Indeed, within a non-Archimedean structure, "infinities"
can relate to one another in uncountably many different ways, and these ways can express
themselves in the usual algebraic operations which are available for them, just as they are
available for the usual real numbers.

\item We miss on the whole of the immensely rich and complex non-Archimedean structure of
"infinitesimals".

\end{itemize}

By the way, in Stochastic Calculus, this nonstandard $^*\mathbb{R}$ is systematically used by
now, ever since 1975, when Peter Loeb introduced his famous nonstandard measure. \\

Needless to say that $^*\mathbb{R}$ is one of the simplest non-Archimedean structures which
has been around. \\ \\

{\bf 5. How Is Mathematics Good for Physics ?} \\

A remarkable and puzzling fact about Mathematics in its relation with Physics and other
sciences was expressed nearly five decades ago by the Physics Nobel Laureate, Eugene Wigner,
[19]. We shall return to that issue later, while in this section we address another, not
totally unrelated aspect of the relationship between Mathematics and Physics. \\

Mathematics, at its best, can be far more than a mere tool at the disposal of Physics. Indeed,
it can be a source of fundamental visions, visions which can lead to new fundamental physical
concepts. And needless to say, that goes for the role of Mathematics not only in Physics.

The mentioned examples of higher, and even infinite dimensional spaces, or of the complex
numbers are some of the most obvious and well known such examples of mathematical concepts
becoming fundamentally important in theoretical Physics. \\

Here the issue is precisely that physicists, hardly ever sufficiently familiar with the best
of the existing Mathematics - except so far with the unique case of Newton, of course - do
inevitably and unknowingly limit their visions. And that goes as well for their visions of the
yet more fundamental realms of Logic, or rather, Mathematical Logic, and not only of
space-time, dimensions, finite versus infinite, and so on. \\

A certain exception to such a customary limitation of vision one can find in [5], where the
amusing question is asked :

\begin{quote}

How come that theoretical Physics has so far never used spaces which have a cardinal larger
than that of the real numbers ?

\end{quote}

Of course, the point is not that one should now by all means start using in Physics spaces
with very large cardinals. However, that question cannot simply be dismissed instantly and
without any thought, given the immense amount of spaces with much larger cardinals, spaces
easily available ever since George Cantor's Set Theory was established, which by now is nearly
150 years old ... \\

By the way, [5] itself is somewhat limited in its view of the possible role of Mathematics in
theoretical Physics, since it is completely missing on the following issues :

\begin{quote}

1. The fact that we have no freedom to choose only Archimedean space-time, [8,9,16,17].

2. The possibility to use inconsistent logic, [4].

3. The possibility to use self-referential logic, [2].

\end{quote}

And now, for brevity, back to the first only of these three issues of our more specific
concerns. \\

In the sequel, we shall make a few comments on the second and third issues above.

Suffice here to mention in this regard that, as well known, Quantum Mechanics is highly
counter-intuitive. And it is so to the extent that it borders on what our usual intuition may
consider to be straight {\it paradoxical}. And then, needless to say, inconsistent logic may
quite likely be one way to explore the study of quantum phenomena.

For that matter, self-referential logic is also known to lead to paradoxes, hence it may have
the credentials to be similarly appropriate for dealing with the quantum world ... \\

The very strong insistence in Smolin's mentioned book on the need for a {\it background
independent} theory is most remarkable indeed. However, what no physicist so far seems to
notice is that it is {\it not} enough to set aside the assumption of a specific, given, fixed,
time independent background. No, it is not at all enough, as long as physicists clearly,
unknowingly and insistently do still {\it hold} so much to the very same {\it nature} of the
structure of the abandoned background, and do so when they envision all those time dependent
backgrounds as being still Archimedean.

And they most certainly {\it do} hold to that vision, that is, the good old Archimedean
one ... \\

And what is wrong with that, when formulated for convenience in a more plain English, and not
in technically involved Mathematical terms ?

Well, in the Archimedean vision there are, and there can only be two kind of entities : \\

{~~~~~~~~~~} finite and infinite. \\

And everything finite can finitely be compared with everything else finite. No finite can be
usefully compared to infinite. All infinite entities are the same kind of infinite. No
infinite can operationally, thus usefully be compared with anything finite. No infinite can
operationally, hence usefully be compared with infinite. \\

Thus the Archimedean vision only knows about {\it two} kind of relationships between the
entities involved, namely : finite, or infinite. \\

And then no wonder that in Smolin's mentioned book, countlessly many times comes up the issue
of "infinities in Physics" ... \\

Yet, for no less than 41 years by now, that is, ever since the 1966 book "Non-Standard
Analysis" of Abraham Robinson, we have one extraordinary successful example of a one
dimensional non-Archimedean structure, namely, the nonstandard real line $^*\mathbb{R}$, in
which simply there are {\it no problems} with any sort of infinities ...

And the reason for that most convenient state of affairs is that, somewhat similar with
Cantor's Set Theory, instead of one single infinity, there is an uncountable hierarchy of them,
and one can do with all of them all of the usual algebraic operations ...

Added to that, what we usually consider finite, turns out to be quite infinite, when compared
with the infinitesimals. And similar with the infinities, there is an uncountable hierarchy
of infinitesimals ...

In short, one infinite can be infinitely larger, or for that matter, infinitely smaller than
another infinite.

And similarly with infinitesimals. \\

The point is that even in that so far most simple one dimensional non-Archimedean case of
$^*\mathbb{R}$, the local and global structure is ... infinitely ... more rich and complex
than what physicists have so far ever dared to envision ... \\

And then, who is there to say a priori that such a thing does not, and can never ever have any
relevance to theoretical Physics ? \\

By the way, $^*\mathbb{R}$ still does not answer the question in [5] about the rather limited
cardinal of spaces used so far in Physics, since the cardinal of $^*\mathbb{R}$ is still the
same with that of the usual real line $\mathbb{R}$. \\

However, in non-Archimedean structures, and even more so with those of higher dimension than
that of the most simple $^*\mathbb{R}$, to be finite is, among others, but a {\it relationship}
between two entities. And so is infinite, or for that matter, infinitesimal. Furthermore, one
can perform with all these relationships all of the usual algebraic operations. And the
immense richness and complexity unleashed by all that is something which should at long last
be considered by certain theoretical physicists ... \\

For instance, in all those higher dimensional spaces of String Theory, how about the unseen
dimensions being in fact of infinitesimal, or for that matter, of infinite size ? \\

That would certainly help in fixing some of those many undetermined constants ? \\

And if we talk about dimensions, then how about non-integer, that is, fractal dimensions ? \\

For more than three decades by now, such dimensions are known and used. Yes, they are used
even in Physics, for instance, in heat propagation, or other diffusion processes, and
consequently, in Probability Theory and Stochastic Processes ... \\

Yes, there is so much more out there which could help the vision of physicists. A vision
hopefully leading to new fundamental concepts ...

And help such vision, much beyond even what is suggested in [5] ...

Just as it happened not such a long time ago, when higher, and even infinite dimensional
spaces were adopted in Physics, and also, the complex numbers ... \\

Amusingly, there are plenty of natural non-Archimedean realms of quite dramatically effective
use, as shown in the references of [7,8,16,17]. And as also shown there, if we accept -
knowingly or not - the Archimedean assumption, we actually condemn ourselves to a very partial
view of things ... \\

By the way of the place and role of Mathematics, and not only related to theoretical
Physics. \\

Whenever I talk in public about Mathematics, that is to physicists, engineers, and so on, I
keep explaining that mathematics, well, is in fact ... {\it not} Mathematics ...

Yes, Mathematics should rather be seen as, so far, the only science devised by us humans which
is both {\it precise} and {\it universal} in its validity and applicability. Indeed, Physics
and Chemistry, for instance, are also quite precise. But they are not as universally valid or
applicable as Mathematics. And the price we pay for this universality of Mathematics is that
it has to be more {\it abstract} than Physics, Chemistry, and the like. \\

So that, so sorry to say, it is rather the fault of the whole of mankind that, for the time
being, we did not develop another, possibly more widely user friendly or easily accessible,
precise and universal science. Especially since most humans seem to dislike precision, and yet
more deeply dislike abstract ideas ... \\

On top of that, we also have of course the following {\it self-reinforcing} effect. Namely,
the fact that Mathematics is not widely user friendly led along the ages to the situation that
Mathematics is only talked about among mathematicians, thus inevitably further increasing its
lack of wider user friendliness ... \\

However, there is - and there has always been - some good news as well. Indeed, Mathematics,
and even more likely any of its possible more user friendly variants which mankind may
eventually manage to develop, is actually {\it not} so impossible for humans at large. And
this fact is thoroughly proven in all human societies. Certainly, nearly all humans, no matter
how unintelligent or illiterate, have not only been most eager to learn the basics of counting,
but even managed it quite well when, for instance, it comes to counting their own money ... \\ \\

After all, Mathematics is mostly about precision and abstraction ...

And having zero amount of money in one's pocket is pretty precise, even if rather unfortunate.
But above all, it is considerably abstract, as proven by the fact that for a long long time,
the number zero did not have a mathematical notation dedicated to it ... \\ \\

{\bf 6. Seers and Craftspeople ...} \\

When I started to read Smolin's mentioned book, I happened to open it at the section "Seers
and Craftspeople" ...

Well, in over fifty years in research, mostly in Mathematics, it was for the first time that I
read such an honest, brave, and above all, highly accurate and relevant account about the
inner working of Physics - an account which unfortunately is quite accurate for Mathematics
and other hard sciences as well - an account written by such a highly reputed scientist ...

But then, the words are less important than pointing to the highly undesirable situation in
present day science. And that situation took a dramatic turn for the worse starting in the
late 1950s, when because of various reasons, among them the sudden fear of the so called
"missile gap", a fear which hit the USA upon the launching of the first soviet sputnik, the
number of scientists was massively increased in the West. \\

Consequently, it was inevitable that relative to those vast numbers, fewer and fewer would in
fact be seers. In addition, the fast growing number of the rest ended up running much of the
show ... \\

So that nowadays, after nearly five decades, we can thank that things are not worse than they
already are ... \\

In this regard, it is so reassuring to see someone of the stature of Smolin stand up and say
to a wide public that there are serious foundational problems in Physics, and among them, in
Quantum Mechanics. Being myself more of an amateur physicist than a professional one, ever
since the late 1950s when I first started to learn the subject, I was shocked to see that even
in its simplest instance of non-relativistic finite quantum systems, the mathematical models
used were far from correct mathematically, [11]. And that situation quite sharply contrasted
with the fact that by that time, such was not at all the case in any of the other more
classical disciplines of Physics, not even in General Relativity. Needless to say, the lack of
mathematical correctness encountered in Quantum Field Theory is still more considerable.
Merely the way the Feynman path integrals are, so called, defined, and then of course used,
not to mention the various manipulations related to renormalization, are nothing short of a
{\it major scandal}, in case they would occur nowadays in more classical branches of
theoretical Physics. \\

Lately, however, I started to find that the foundational issues reach deeper than usually
thought, and in fact they may affect just about all the present mathematical models of much of
Physics, including those which are background free.

For brevity, let me repeat that the respective deeper foundational issue is a consequence,
among others, of the assumption that space-time is Archimedean, an assumption which we got
stuck with ever since ancient times, an in particular, since Euclid.

It may also be, as mentioned, a consequence of the blind belief in the validity of the scaling
group, namely that its structure does hold on arbitrary macro-scales, although in view of the
atomic structure of matter, we already know very well that the same is not the case on
micro-scales. \\

Amusingly, the situation is further aggravated by the general perception that, in fact, we do
have the freedom to chose between the Archimedean and non-Archimedean assumptions, and that
our choice of the Archimedean one is therefore a free choice, and one based on suitable
arguments.

In case the non-Archimedean alternative for space and time is considered, an extraordinary
rich and complex structure follows which, recalling fractals among others, has a {\it
self-similar} structure. \\

This self-similarity in the case of time means among others that each instant does in fact
contain an uncountable amount of {\it eternities}, and that {\it beyond} what we usually
consider eternity, there are uncountable other eternities, both at the "beginning" of time,
and at the "end" of it.

Correspondingly, as argued in [7,8,16,17], we do not actually have a freedom of choice, since
the Archimedean assumption automatically locks us up into "one single walkable world". And in
such a situation we impose upon ourselves the fact that whatever is infinite exists only
faraway at infinity, while everywhere all around us finiteness prevails.
In this way, we do not and cannot encounter eternity in the "now", and we can only know about
at most two eternities, namely, one before the "beginning" of time, and one after the "end" of
time ...

But then, we may not even have that ... modest wealth ... of eternities, since according to a
rigorous interpretation of the Archimedean assumption on the structure of time, there cannot
be anything  before the "beginning" of time, or after the "end" of time. Instead, we only can
have one single eternity, namely, reaching unlimited in the past, and in the future ... \\

On the other hand, what is both amusing and important to note is that non-Archimedean
structures have most successfully been used since the mid 1960s in obtaining generalized
solution for very large classes of linear and nonlinear partial differential equations, see
[18] and the references in [7,8,16,17], as well as 46F30 at \\

{~~~~~~~~~~} http://www.ams.org/msc/46Fxx.html \\

for the whole respective subject in Mathematics. \\

Consequently, the idea of using non-Archimedean mathematical models in theoretical Physics
should not be so easily dismissed as a mere fancy type mathematical idea. \\

Well, one may simply say that abstract means among others non-physical ...

After all, "abstract" is about ideas, Platonic or not, while Physics is, well, about ...
Physics ... \\

And then, what started to disturb me lately with respect to the mathematical modelling of
Physics is that what appear to be exclusively physical reasons, and at that, most elementary
and primitive ancient ones, we ended up being so completely stuck into an Archimedean
perception of space and time, and in fact, of much else which is quantitative.
No wonder that we hit "infinities in physics", be it in General Relativity or Quantum Field
Theory, or for that matter, even in such classical realms as shocks or turbulence ... \\

Of course, the p-adic story, which happens to be non-Archimedean, has been around for quite a
while, and it is pretty clearly established and with lots of very good results. \\

So is Nonstandard Analysis ... \\

Yet what makes the story without much impact among physicists - even among those few who are
seers - is the mistaken illusion that, just as with other axioms or assumptions, we have the
total freedom to chose between the Archimedean and non-Archimedean assumptions. And then of
course we choose the former, since obviously it is so much simpler in itself and in its
consequences ... \\

Furthermore, even those mathematicians who deal with non-Archimedean structures, like for
instance those in Nonstandard Analysis, completely fail to realize that we do not have the
above mentioned freedom of choice. But then of course they are far too much focused on what
can be transferred and what cannot from usual Analysis into Nonstandard Analysis. So much so
that they completely miss to note, let alone use, the surprisingly rich and complex
self-similar structure even of the nonstandard real line $^*\mathbb{R}$. \\

Consequently, what matters is that the Archimedean choice quite hopelessly limits our intuition
and vision, and keeps doing so for millennia by now, without us ever being aware of that ... \\

And then perhaps, modern physics may by now really need much much more ... \\

Recently a few physicists have given some thought to the possible role in physics of scalars
other than the usual real or complex numbers.

In this regard, the use of octonions is the farthest one can go along the ... good old
classical ... lines of mathematical thought. And they are quite troublesome, since they fail
not only to be commutative, but even associative. \\

On the other hand, as it happens, there infinitely many other easily available scalar systems
to be used, see [8,9,16-18] and the references therein ... \\

The three fundamental aspects such scalar systems have are :

\begin{quote}

1) They are non-Archimedean

2) They are not fields, but only algebras.

3) They have during the last four decades proved to be invaluable in solving very large
classes of earlier unsolved nonlinear partial differential equations, as seen at \\

www.ams.org/msc/46F30

\end{quote}

It was with Newton last time, and quite likely the first time as well, when a major revolution
was made in Physics and it was the respective physicist who all by himself created the needed
Mathematics, namely, Differential and Integral Calculus. \\

When Einstein brought about Special Relativity, the Mathematics needed was of an elementary
school level, and he happened to know it. But when he went on to General Relativity, he did
not in the least make the respective Mathematics, that is, Differential Manifold Theory, and
instead, he had to rely on his mathematician friends. \\

Well, not much later, Quantum Mechanics, as established in the 1920s by the respective
physicists, did not use any new Mathematics. And it was "put right" mathematically - with all
the troubles we now know about - by von Neumann. \\

Nowadays however, physicists do not in the least think that in truly foundational issues they
should imitate Newton. And amusingly, as far as their view of Mathematics is concerned, they do
not even think that they should perhaps try to imitate Einstein when he was bringing forth
General Relativity. \\

As for mathematicians, hardly any of the really good ones, or of those with really good new
ideas, are concerned about Physics to any practically relevant, let alone effective degree. \\ \\

{\bf 7. Two Further Fundamental Mathematical Ideas \\
        \hspace*{0.5cm} Physicists Have ... No Idea About ... } \\

Amusingly, for some time by now, two absolutely revolutionary ideas have been introduce in
Mathematics, even if hardly anyone knows about them within that very discipline itself. Namely :

\begin{quote}

1) Self-referential logic, [2].

2) Contradictory logic, [4].

\end{quote}

As for 1), ever since the Russell paradox, or in fact, of its ancient Greek version of the
paradox of the liar, we have quite dramatically avoided any kind of self-referential
statements in Logic or Mathematics.

Well, some years ago, very serious people started to develop a very serious theory of such
earlier ... totally forbidden ... kind of statements. A good account of the respective
developments is presented in [2]. \\

About 2), there have of course been similar most grave concerns ... \\

On the other hand, we all use our digital computers which, as far as their computation of
integers is concerned, function according to :

\begin{quote}

2.1) the Peano axioms,

\end{quote}

plus the axiom

\begin{quote}

2.2) there exists $M  >> 1$, such that $M + 1 = M$

\end{quote}

where of course $M$ may be $10^{1000}$, which is the respective "computer infinity". \\

And obviously, 2.1) and 2.2) constitute a {\it contradictory axiomatic system} ! \\

And yet, we spend money on buying such computers, and when we fly on our planes, we may be
afraid of hijackers, but certainly not of the computers used to design those planes ... \\

Amusingly again, some years ago, serious people started to develop a logic of contradictory
axiomatic systems. An account of such developments can be found in [4]. \\

And who can a priori and competently say that, when dealing with modern foundational issues of
Physics, one would not have to go so far as the two ideas mentioned above ? \\ \\

{\bf 8. Is Physics to Remain Physics ?} \\

The recently emerged claim - as suggested among others by the massive development of the
theory of Quantum Computation, and strongly supported by a variety of physicists - that
"information is physical", offers the opportunity for a better look at what may in fact be
involved. Here, in view of the ideas presented above, we shall venture several concluding
comments. \\

As already mentioned, and as should be clear upon a minimally careful thinking, the formula
"information is physical" cannot make much sense unless both of its terms are defined to a
satisfactory extent. Regarding {\it information}, and when considered in the context of
present day Information Technology, its definition is quite clear, and since Shannon, it even
has a rigorous way for being measured. \\

But can one say quite the same about the term {\it physical} ? \\

Here what happens is rather that we face a certain kind of vicious circle which manifests
itself in several ways.

First, when one asks a physicist what may the term {\it physical} mean, one may typically
encounter a reaction which says that the term is of course self-evident to any physicist,
while those for whom it is not, only show that they are outside of Physics ...

In other words, according to such physicists, physical is supposed to be what physicists are
interested in and busy with ... \\

A somewhat more benevolent, edifying, and less of a vicious circle type reaction may go along
the lines of a certain attempt at a popularizing kind of explanation of the meaning of the
term, and of course, of the realms which it is supposed to refer to.

Yet here as well, some kind of vicious circle still remains due to the fact that the given
explanation does, and in fact must quite inevitably use physical concepts and terms ... \\

But then, we should not hold that against the physicists. Indeed, quite likely the same must
inevitably happen when a specialist in any other field of science, including Mathematics, for
instance, may try to give a definition of the respective field ... \\

And then, what may the problem be with that latest formula "information is physical" ? \\

Well, for one, and as mentioned above, the concept of {\it information} has not yet entered
among the fundamental concepts of modern Physics, and certainly, it does not yet have a place
of equal importance with the stone age type fundamental concepts of position, momentum, mass,
energy, etc.

And then, is the latest insistence on the formula "information is physical" but a less than
conscious, less than explicit expression of a deeper awareness and desire at last to enlarge
that stone age - and engraved in stone - collection of fundamental physical concepts ?

And in case that may indeed be the underlying story, then we can only wonder why that formula
attempts to do so merely by {\it reducing} information to the physical, to that very same
stone age one ? \\

Here therefore, we can note yet another most strong - and so far, awfully successful - attempt
of a long established paradigm to survive regardless of everything ...

Yes, Physics wants by all means to remain Physics ...

Even if fundamental new concepts must in a rather dubious manner be seen as simply reducible
to the good old ones ... \\

Amusingly, in the late 1800s, Mathematics had for a while gone through a similar stage. Indeed,
most influential mathematicians at the time, among them a true giant of the field like Henry
Poincar\'{e}, were irrevocably against the introduction by Georg Cantor of the concept of {\it
set}, and of the respective Set Theory. Yet by the 1920s, all of Mathematics started to be
reformulated in terms of Set Theory, and by the 1950s, that project was accomplished.

And ever since, modern Mathematics can in a way be seen but as a kind of specialized branch of
Set Theory ... \\

More amusingly still, in the 1940s, while the reformulation of all of modern Mathematics in
terms of Set Theory was going on, a yet more fundamental and general mathematical theory was
introduced, namely, Category Theory.

In this regard, the extent of the success of Set Theory as foundational for modern Mathematics
can be seen, among others, in the fact that there has not yet been a similar reformulation in
terms of Category Theory.

Needless to say, there is also another reason for that delay. Namely, as in all human ventures,
so in Mathematics, established paradigms tend to have a strong staying power. And the
revolution brought about by Category Theory may actually have come too soon after that of Set
Theory. Too soon to tempt many enough mathematicians away form the remarkably successful
workings of modern Mathematics as formulated in terms of Set Theory. Also, Category Theory
happens to be significantly more abstract and involved than Set Theory. Therefore, the vast
majority of the so called "working mathematicians" - which of course is merely another term
for Smolin's "craftspeople" - do not feel the need to do the respective investment of time and
effort in order to switch from Set Theory to Category Theory. \\

And when mentioning here these revolutionary events in the foundations of modern Mathematics,
one should not miss the opportunity that no less revolutionary events happened during the last
century and half in Mathematical Logic. One of them, the emergence in the 1950s of Model
Theory, led in the 1960s to Nonstandard Analysis, as introduced By Abraham Robinson. \\

And still, after all these revolutions, Mathematics is still Mathematics, and Mathematical
Logic is still Mathematical Logic ...

Yet both these disciplines, with respect to their fundamental concepts, are indescribably
beyond the stages they happened to be less than two centuries ago ... \\

And then, what is that which keeps Physics, while remaining still Physics, from undergoing
similar revolutions in its most fundamental concepts ?

Revolutions, which may at last see it past the stone age cage commented upon above ? \\ \\

{\bf 9. Deeper than Physics and Mathematics ?} \\

\begin{quote}

\begin{quote}

\begin{quote}

"When the province of physical theory was extended to encompass microscopic phenomena through
the creation of quantum mechanics, the concept of consciousness came to the fore again. It was
not possible to formulate the laws of quantum mechanics in a fully consistent way without
reference to the consciousness." \\

\hspace*{1cm} E Wigner, Nobel Laureate in Physics \\ \\ \\

"Nevertheless, the physics community does not accept the study of consciousness itself as part
of our discipline. And that is appropriate. Consciousness is too ill-defined, too
emotion-laden. It is not the sort of thing we deal with in physics. But discussion relating
quantum mechanics and consciousness will not go away." \\

\hspace*{1cm}                                B Rosenblum, F Kuttner :

\hspace*{1cm}                                Quantum Enigma (pp. 4,5)

\end{quote}

\end{quote}

\end{quote}

Wigner's paper [19] had at its time elicited a number of comments. As it happened however,
they did not seem to go deep enough in searching for the {\it reasons} of that truly
remarkable "unreasonable effectiveness of Mathematics in the Natural Sciences" ... \\

The reason for that failure seems quite simple and obvious. \\

The moment some professional philosophers may become involved in such, or for that matter, any
other possible discussions regarding the deeper meaning and impact of Mathematics, they cannot
help but bring into play their typical unlimited freedom in points of view. And also quite
typically, few, if any of such points of view may gain the agreement of so called "working
mathematicians", or in Smolin's terms, "craftspeople" in Mathematics.

After all, philosophers cannot help taking the "bird's eye view", and not seldom that of a
"bird" which may have somehow managed to fly off the specific "ground" that originated the
discussions ...

Consequently, its is unlikely that either mathematicians or philosophers may really manage to
benefit, since the discussions remain of interest only to some philosophers ...

In this way, the respective discussions are quite likely to end after some time, and do so
without much relevance ... \\

It may as well happen, as was actually the case with Wigner's mentioned paper, that a few
"working mathematicians" get involved in the discussions ...

In such a case, however, a certain symmetric effect tends to happen, and does so with few
exceptions, if at all.

Namely, "working mathematicians" tend to take the "worm's eye view" of the issue. And such a
view is equally likely to lead to irrelevance ... \\

But then, there may be a third trouble as well, and why not, a corresponding {\it third
way} ...

And if we happen to talk about various ways of viewing, then a most appropriate analogy is
obtained by recalling that :

\begin{quote}

We see through our own eyes, yet in the process of seeing, we do {\it not} see our own eyes,
unless there is something wrong with them ...

\end{quote}

Well, when doing Mathematics, or for that matter, any other science, we are of course {\it
thinking}. And yet, we take that thinking process so much for granted that we hardly ever stop
even for a moment to {\it think about it} ...

And certainly, no "working mathematician" worthy of that name, or for that matter,
"craftspeople" in Physics, would ever think of doing so ... \\

And the consequence ? \\

Well, some of the consequences were mentioned in [7]. Here we recall them briefly. \\

For instance, within Newtonian Mechanics, both space and time are absolute. And as such, they
are supposed to contain all that exists, including the bodies of the thinking scientists.

What however is less clear is whether Newtonian space-time is as well supposed to include the
very thinking of the respective scientists ... \\

As for Einsteinian Mechanics, with its severe limitation on the velocity of propagation of all
sorts of physical phenomena, the fact nevertheless remains that just about every human, no
matter how incompetent in Physics, and in particular, in Special or General Relativity, can
quite easily think at the very same moment about two different places in the universe, no
matter how faraway those two places would happen to be from one another ...

Thus such a thinking simply does not conform to absolutely any limitation ... \\

In Quantum Mechanics we encounter as similar situation. Indeed, two entangled particles A and
B may perhaps not be able to communicate instantly with one another the observed state of one
of them.

And yet, anyone familiar enough with Quantum Mechanics, can perfectly understand in an instant
what happens with two such entangled particles, no matter how far they would be from one
another ... \\

And then, when trying to answer the question :

\begin{quote}

"where and how does all of that thinking happen ?"

\end{quote}

we do face a question which modern science, and specifically Physics, not only has no answer
to, but in fact, it simply does not care about. \\

Or rather, it does not care to care about ... \\

And which of the modern sciences should care about it ? \\

After all, all the processes involved in the examples mentioned above, or in [7], are clearly
"physical" in their nature. Thus so sorry to say, from all sciences, they are nearest to
Physics ... \\

Of course, if anyone cares about such a question, there may be quite a number of avenues to
pursue, and needless to say, not all of them may be proper ... \\

Here, we shall only mention two of them, and leave it to the reader to consider to which
extent they may happen to be appropriate ... \\

First, we could quite easily take a page from Descartes, and in terms of his celebrated "res
cogitans" and "res extensa", say that, of course, the answer to the above question must happen
nowhere else but in "res cogitans" ... \\

However, by doing so, we better start by noting that it has for long been fashionable to label
Descartes a "dualist", or even more vulgarly, a "substance dualist".

What is missed in such a judgement is the simplest understanding of the world-views of
thinkers in the Europe of those times. To mention a few of them, Copernicus, Kepler, Galileo,
Pascal, Descartes, Newton, Leibniz, or Spinoza were deeply religious men in the
Judaeo-Christian tradition. Consequently, none of them - and this includes Descartes as well -
could possibly be anything else but fervent "monists". \\

As for "dualism", or for that matter, "substance dualism", Chemistry is practicing it without
any objections from any quarter, and it does so in a most successful manner, when it divides
itself into its "inorganic" and "organic" branches. \\

Biology does the same when it makes an essential differentiation between "living organisms"
and all other forms of matter. And such a differentiation is by no means arbitrary or
superficial. For instance, only plants are able to turn through their metabolism inorganic,
thus clearly non-living matter, into living one. And by far most of the plants only use
inorganic matter in their metabolic processes. Animals, on the other hand, must use in their
metabolism mostly plants or other animals, since they cannot live only on inorganic intake. \\

Regarding Descartes, his division in "res cogitans" and "res extensa" was of course but about
the two branches of a tree which grow out from the same one and only, universal and all
encompassing, eternal grace of God's act of creation. \\

As for modern Physics, there appear to be two rather different ways "res cogitans" and "res
extensa" happen to relate to one another. \\

In pre-quantum Physics, including General Relativity, the respective theories are of course in
"res cogitans". And their setting up, as well as testing, can be done in ways which do {\it
not} interfere with "res extensa", where the actual physical phenomena studied take place. In
short, by looking at the Moon, for instance, and doing so with one's naked eye, one is not
supposed in any way to affect the motion of the Moon. \\

On the contrary, the Copenhagen Interpretation of Quantum Mechanics has as one of the basic
axioms the so called "collapse of the wave function". This leads, among others, to the
celebrated paradox of "Schr\"{o}dinger's cat", according to which, the simple fact of looking
with one's naked eye at the content of the box in which the respective cat was placed is
supposed to make all the difference for that poor creature between its life or death, since it
"collapses" the corresponding wave function. In this way, here we are supposed to have a much
different relationship between "res cogitans" and "res extensa", when compared to that in
pre-quantum Physics. \\

As it happens, this different and novel type of relationship between "res cogitans" and "res
extensa" has led to a variety of suggestions and speculations, some of them possibly being
exaggerated.

Here we can recall among the better known ones "Wigner's friend", "quantum suicide", or why
not, even "quantum immortality" ... \\

By the way, we can note as well a related omission in the interpretation of the celebrated EPR
experiment.

Let us assume that the two entangled quantum particles A and B in the EPR experiment are such
that whenever one has the spin "up", the other must have the spin "down". If now an observer P
situated at A measures the spin of A and finds it "up", then this observer can instantly know
that the spin of B - no matter how faraway - must be "down", and vice-versa.

Of course, if an observer Q is placed at B, then P is not supposed to be able to {\it
communicate} with Q instantly what the spin of B is. \\

And yet P can instantly {\it know} what the spin of B is, as soon as P measures the spin of
A. \\

Clearly, in view of Special or General Relativity, this instant {\it knowledge} by P of the
spins of both A and B cannot take place in "res extensa", and instead, it is rather happening
in the "res cogitans" ... \\

But let us say now that this instant knowledge does not take place either in "res extensa", or
in "res cogitans" ...

Then perhaps on a nice day, some "seer" type physicist may become curious, and find a third
realm beyond the two Cartesian ones, {\it where} that instant knowledge happens ... \\

Moreover, once our "seer" type physicists finds such a third realm, he or she may become
curios and interested about {\it how} such an instant thinking happens ? \\

After all, if by now it is so loudly claimed by many physicists that "information is physical",
then what is wrong with taking one more step and considering that "thinking is also
physical" ? \\

Would not such one mere more step add incomparably to the glorious march of Physics on its way
to keep enlarging its realms of interest for evermore ? \\

Anyhow, in case taking a page from Descartes may happen not to be so tempting for some, then
perhaps, we can go way back in time, to the ancient Greeks, and the {\it Paradox of the
Liar}. \\

Formulated in one of its simplest forms, it is given by the sentence :

\begin{quote}

"This sentence is false."

\end{quote}

In modern times, this paradox obtained quite some importance. In the early 1900s, Bertrand
Russell reformulated it in terms of Set Theory, thus precipitating for a time a massive
interest in the Foundations of Mathematics.

In the early 1930s, Kurt G\"{o}del used it as a basic idea in proving his two celebrated
Incompleteness Theorems. \\

As for modern attempts at the explanation of that ancient paradox, one of the basic ideas has
been the essential distinction in Semantics between a {\it language} and its {\it
meta-language}. And in view of such an explanation, the trouble with the above paradoxical
sentence is simply in the fact that, in an inadmissible manner, it mixes up these two distinct
levels of language. \\

However, what is of interest to us here is not so much Semantics, or the respective
explanation of that ancient paradox.

After all, the issues raised by that paradox are far from simple. And even in its much more
specific context of Set Theory, it did lead to at least three very different ways which tried
to explain it and overcome it, namely, Logicism, Formalism and Intuitionism. \\

Instead, what we may easily note is its essential and unbreakable linking of a "statement"
with its "interpretation", a linking in an endlessly ongoing cycle. \\

And quite clearly, semantics or no semantics, there is some undeniable {\it difference}
between a statement and its interpretation ... \\

Therefore, a message of that ancient paradox is simply the following :

\begin{itemize}

\item Either we like it or not, there are at least two rather different realms the moment we
start to speak, and hopefully, prior to that, to think.

\item And a careless dealing with these two realms, let alone, the disregard of the difference
between them, can so easily lead to trouble.

\end{itemize}

And this essential difference between stating and interpreting need not lead either to
Semantics, or to the Cartesian "res extensa" and "res cogitans". Also, it need not imply the
complete disjointness of the realms of stating from the realms of interpreting. Furthermore, as
shown abundantly in [2], it need not lead to the instant disqualification of self-referential
statements. \\

Instead, the moment we reach a situation where thinking becomes involved in certain
paradoxical situations involving an essential linkage between two or more realms, as already
happens in Physics, and it has done so at least since Special Relativity, see [7] and the
above related brief comments, we should no longer simply keep dismissing the situation. Nor
should we merely try to avoid it by focusing exclusively on one of the realms involved, the
so called "physical" as happens to be understood by the Physics of our present time, and
leaving the other realm, or possible realms, to be deal with, if at all, by anybody
else ... \\

Naive Set Theory, as it had been developed prior to the emergence of paradoxes such as that of
Russell, had it great successes ...

Yet it had to be left behind in favour of its more deep and systematic development  ... \\

Is present day Physics in a similar "naive" stage, when avoiding even to consider, let alone
answer, the above type question of :

\begin{quote}

"where and how does it happen ?"

\end{quote}

\end{document}